\pdfoutput=1
\documentclass[a4paper, 11pt]{article}
\usepackage[margin=2.4cm]{geometry}
\usepackage{xcolor}
\usepackage{graphicx}
\usepackage{listings}
\usepackage{amsmath}
\usepackage[natbibapa]{apacite}
\usepackage{hyperref}

\lstdefinestyle{numbers}{language=R,%
  basicstyle=\ttfamily\small,%
  commentstyle=\slshape\color{blue!50!black},%
  numbers=left,numberstyle=\tiny\color{gray!40!black},%
  basewidth={.5em, .4em},%
  showstringspaces=false}

\newcommand{\mytab}[2]{%
\begin{tabular}{@{}p{0.15\textwidth}p{0.85\textwidth}@{}}%
#1 & \parbox[t]{\linewidth}{#2}%
\end{tabular}%
}

\newcommand{\vect}[1]{\mathbf{#1}}
\newcommand{\mat}[1]{\mathbf{#1}}
\newcommand{\gvect}[1]{\boldsymbol{#1}}
\newcommand{\gmat}[1]{\boldsymbol{#1}}

\title{\vspace*{-35pt}Simulating the Power of Statistical Tests:\\
  A Collection of R Examples}
\author{Florian Wickelmaier\thanks{Florian Wickelmaier,
  Department of Psychology, University of Tuebingen,
  Schleichstr.~4, 72076 Tuebingen, Germany,
  \texttt{florian.wickelmaier@uni-tuebingen.de}}\\
  \emph{University of Tuebingen}}
\date{}

\begin{document}
\maketitle

\begin{abstract}
\noindent
This paper illustrates how to calculate the power of a statistical test by
computer simulation.  It provides R code for power simulations of several
classical inference procedures including one- and two-sample $t$ tests,
chi-squared tests, regression, and analysis of variance.
\end{abstract}
   
\section*{Introduction}

\subsection*{Statistical power and inference from statistical tests}

Classical (also called frequentist, as an alternative to Bayesian) statistics
provides a framework for deciding between two hypotheses about one or more
unknown parameters of a model that generated the data \citep{NeymanPearson33}.
The null hypothesis (H$_0$) has to be specific such that the data-generating
model is fully specified or only depends on quantities that can be estimated
from the data. For example, in a binomial model the success probability is
fixed to, say, H$_0$: $\pi = \frac{2}{3}$. The alternative hypothesis (H$_1$)
is the logical opposite of H$_0$, for example $\pi \neq \frac{2}{3}$. Since
the true parameter value is unknown, the decision can either be correct or
wrong. Choosing H$_1$ when H$_0$ is true is called a type~I error, its
probability $\alpha$ is usually set to $0.05$. Choosing H$_0$ when H$_1$ is
true is a type~II error, its probability $\beta$ is often set to be $0.2$ or
lower.

In order to decide between H$_0$ and H$_1$, data will be collected and the
test statistic and its p-value calculated. The p-value is the probability of
obtaining a test statistic that signals a deviation from H$_0$ at least as
extreme as that observed in the experiment, given H$_0$ is true and its
underlying model holds. The decision rule is that if a test is significant so
that its p-value is smaller than $\alpha$, then H$_1$ is chosen, else H$_0$ is
retained.

An experiment should be designed so that it has a good chance to detect an
interesting deviation from H$_0$. Such a deviation is called the effect size.
The sample size of the experiment, $\alpha$, $\beta$, and the effect size are
related. Fixing three of them determines the forth one \citep{Cohen92}. Power
(also called true positive rate, hit rate, sensitivity, or recall) is defined
as $1 - \beta$. It is the probability of a statistical test to detect an
effect of a given size. Therefore, designing an experiment to have a good
chance to find an effect means making sure its power is high enough.

\subsection*{High power is a necessary condition for valid inference}

Why is high power so important for a study? Or conversely, why is low power so
bad? A manifestation of the adverse consequences of low power is the
replication crisis in science \citep{Gelman16, OSC15Science, Ioannidis05}.
Hardly will effects be found again if they resulted from underpowered studies.
It was long understood that low power leads to many non-significant tests and
thus to uninformative studies \citep{Cohen90}. What was less obvious, however,
is that in those cases when the test actually is significant, it gets even
worse. What may look like a lucky shot -- one has managed to reach
significance against the odds, so seemingly, the effect must be really large
-- is a statistical artifact. If an underpowered test is significant, the
magnitude of the effect will be exaggerated, sometimes strongly \citep[type M
error;][]{ButtonIoannidis13, GelmanCarlin14, VasishthGelman21}. This is so,
because with low power only the bizarrely extreme effect estimates will make
it across the significance threshold \citep[for example][]{GelmanCarlin14,
Gelman20}. And, still worse, the probability of the effect estimate having the
wrong sign becomes high \citep[type S error;][]{Gelman14}. This is why high
power is a necessary condition for valid inference from statistical tests.
Failing to meet this condition renders misleading results.

\subsection*{Calculating power by simulation}

How does one assure that the power of an experiment is high enough? And, first
of all, how does one calculate the power of a planned test? This guide
addresses both questions by computer simulation. But why calculate power by
self-made simulation rather than by readily available software? Simulation
forces us to specify a data model and to attach meaning to its components.
This in turn will facilitate drawing valid inferences from the statistical
test. After all, it is the model and its implications that the test confronts
with empirical observations. Understanding what exactly is rejected if the
test becomes significant, will improve our conclusions. The idea with power
simulation is to (1) generate observations from a model that includes the
effect of interest and (2) use these observations to test H$_0$. Repeating
data generation and testing, one calculates the power of the test from the
proportion of significant results. In more detail, the steps required for
calculating power by simulation become:

\begin{enumerate}
\item Specify the model including the effect of interest.

This is the most crucial step, and it consists of several parts. First, choose
the statistical model. For a parametric test, its assumptions predefine the
distribution from which observations are drawn. For example, a binomial or a
$t$ test requires the data to be binomially or normally distributed. Second,
fix unknown quantities. In normal models, this requires fixing one or more
standard deviations. Plausible values can be taken from the literature about
broadly similar studies. One caveat is that reported standard deviations often
are underestimates because of outlier removal and because focusing on
significance filters out high values. Third, specify the effect of interest.
Importantly, this is not the true effect or the effect one expects or hopes to
find. One just does not know ahead of the experiment how large the effect will
be; else one would not run the experiment in the first place. Also, it should
never be an effect size taken from another study. Because of the significance
filter, reported effect sizes are almost certainly an overestimate
\citep{Gelman11}. Rather, it should be the biologically or clinically or
psychologically ``relevant effect one would regret missing''
\citep{Harrell20}.

\item Generate observations from the model.

Once the data-generating model is fully specified, it can be used to draw
observations from it. In R, functions like \texttt{rbinom()} or
\texttt{rnorm()} will accomplish this. The number of draws equals the planned
sample size of the experiment. Care has to be taken that the random sampling
from the model reflects the design of the study. If there are dependencies in
the data, as there are with repeated measures, the model has to produce
correlated observations (for an example, see \ref{sec:pairtwosamplet}). These
correlations have to be pre-specified as well.

\item Test H$_0$.

With the model-generated observations at hand one performs a test of H$_0$.
The test should be the same as that which will be used to analyze the actual
data, or else the power of the interesting test will remain unknown.

\item Repeat.

Repeating the data generation and the testing steps will produce a number of
test results, one for each repetition. Power is then the proportion of times
where the test actually gave a significant result.
\end{enumerate}

In pseudo code, these steps become:
\begin{lstlisting}[style=numbers]
  Set sample size to n
  replicate
  {
    Draw sample of size n from model with minimum relevant effect
    Test null hypothesis
  }
  Determine proportion of significant results
\end{lstlisting}

How is this procedure used for sample size calculation? This is
straightforward. One starts with an arbitrary sample size (number of draws)
and calculates the resulting power. If power is too low, one increases the
sample size. This continues until the desired power level is reached. The goal
here is not to find a single seemingly exact number. There are just too many
unknowns: The model never matches exactly the true data-generating process,
the standard deviations and correlations used in the simulation are only
approximate, and more likely than not will there be invalid observations that
have to be excluded due to reasons difficult to anticipate. Thus, for a
realistic study, one should allow for some leeway by assuming higher
variation, less (or more) correlation, and possibly smaller interesting
effects. Yet, power simulation sets the frame of reference for interpreting
the results. The question it answers is: What if? What results can we expect
from a statistical test if the specified set of assumptions about the
data-generating process were true?

\subsection*{An example: Birth rates of boys and girls}

As an illustration of the power simulations listed below consider an
application of a binomial test to study birth rates of boys and girls. To
provide some context, let us assume an experiment is planned to test whether
the probability $\pi$ of a boy birth ($X = 1$) is slightly higher than that of
a girl birth ($X = 0$). The deviation from $\pi = 0.5$ that is biologically
just relevant so that we would regret missing it, is $0.015$. What is the
sample size $n$ required to detect such a deviation?

The assumption implied by the binomial test is that the entire sample is
independent and identically binomially distributed, formally $X_1, \ldots, X_n
\sim B(1, \pi)$ i.i.d. The hypothesis to be tested is H$_0$: $\pi = 0.5$. The
following R code performs a power simulation under these assumptions:

\begin{lstlisting}[style=numbers]
  n <- 9000                                 # sample size
  pval <- replicate(5000, {                 # replications of experiment
    x <- rbinom(1, size = n,                # data-generating model with
                prob = 0.5 + 0.015)         #   minimum relevant effect
    binom.test(x, n = n, p = 0.5)$p.value   # p-value of test against H0
  })
  mean(pval < 0.05)                         # simulated power at alpha = 0.05
\end{lstlisting}

First, the sample size is set to an arbitrary value. Next, using
\texttt{replicate()} the experiment as described above is repeated 5000 times.
Each time, binomial data are generated assuming that observing a boy is
slightly more likely than a girl and, using these data, H$_0$ is tested. The
vector \texttt{pval} contains the p-values resulting from the 5000
experiments. Final step is to count how often the p-value is smaller than
$\alpha$, and thus significant. Here, the proportion of significant results is
about 0.8. In conclusion, the sample size of $n = 9000$ is sufficient to
detect a deviation of $0.015$ with a power of 0.8.

\subsection*{Power function and power curves}

So far, we have discussed how to obtain a single power estimate for a specific
combination of values. Often, it is more informative to show the resulting
power for an array of different settings. In order to do so, we consider the
power of a test as a function of its influencing variables, such as effect
size and sample size. With the power function at hand, we can call it for all
the settings of interest and graphically display the results. Such power
curves visualize the operating characteristics of the test. To return to the
birth-rates example from above, the power function for the binomial test is
defined as follows:

\begin{lstlisting}[style=numbers]
  pwrfun <- function(d, n, alpha = 0.05) {      # define power function
    pval <- replicate(5000, {
      x <- rbinom(1, size = n, prob = 0.5 + d)
      binom.test(x, n = n, p = 0.5)$p.value
    })
    mean(pval < alpha)                          # return power estimate
  }
\end{lstlisting}

\begin{figure}
\centering
\includegraphics[scale=0.75]{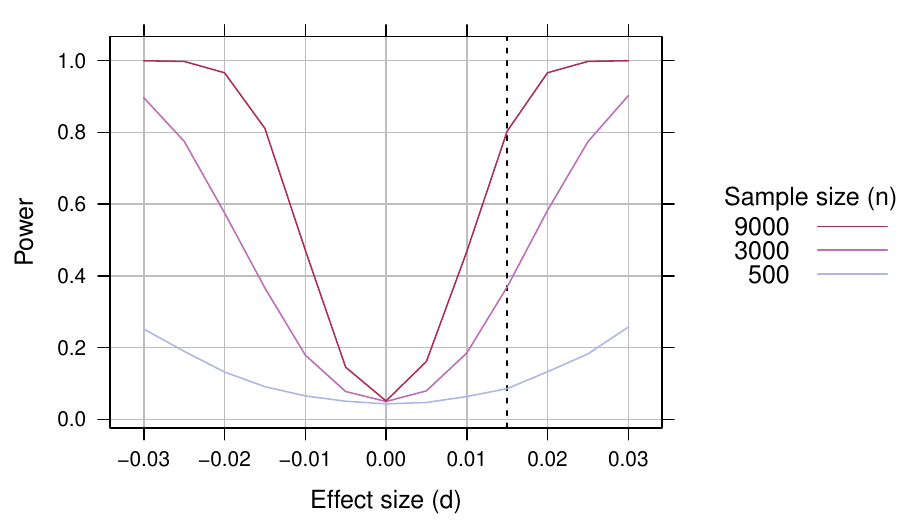}
\caption{Power of the binomial test in the birth-rates example (H$_0$: $\pi =
0.5$) as a function of effect size and sample size. The dashed line marks the
target effect of $d = 0.015$ where power hits about $0.8$ for $n =
9000$.\label{fig:OCbinomtest}}
\end{figure}

The function can be called to produce the power estimate for a certain
setting, for example, \texttt{pwrfun(d = 0.015, n = 9000)} will recalculate
the result from before. When all the settings of interest are combined in a
conditions data frame, they are passed to the power function like this:

\begin{lstlisting}[style=numbers]
  cond <- expand.grid(d = seq(-0.03, 0.03, 0.005),    # effect sizes
                      n = c(500, 3000, 9000))         # sample sizes
  cond$pwr <- mapply(pwrfun, d = cond$d, n = cond$n)  # call power fct. on all settings
\end{lstlisting}

This will call the power function on every combination of values one at a
time. Figure~\ref{fig:OCbinomtest} shows the resulting power curves. Power
rises from $\alpha$ with increasing deviation from $\pi = 0.5$ and with
increasing sample size.

\section*{R code for power simulations of specific tests}

The remainder of this text lists a collection of R examples for power
simulations of specific statistical tests. An overview of the tests covered in
this collection is given below. For each test, there will be a brief
description of a possible application context, of the model and/or its
assumptions, and a statement of H$_0$. For all examples, $\alpha$ is set to
$0.05$. All tests are two-sided. The sample size is chosen so that the
resulting power is about $0.8$. In order to check whether the model has been
correctly translated to R code, a parameter recovery study prior to the power
simulation may be helpful (for example, see \ref{sec:multireg}). It is assumed
that the reader is familiar with the tests listed here. Most of them are
covered in introductory textbooks \citep[for example,][]{AgrestiFranklin21}.

A note of caution: The specific numbers (for effect sizes, standard
deviations, or correlations) do not easily generalize to other application
scenarios. Setting these parameters to plausible values requires
substance-matter knowledge and poses a new challenge for every power analysis.

\tableofcontents

\section{One-sample tests}

\subsection{Binomial test}\label{sec:binomtest}

\mytab{Context}{An experiment is planned to test whether the probability $\pi$
of a boy birth ($X = 1$) is slightly higher than that of a girl birth ($X =
0$). Goal is to detect a deviation from $\pi = 0.5$ by $0.015$.}

\mytab{Assumptions}
{1.\ Exact test: $X_1, \ldots, X_n \sim B(1, \pi)$ i.i.d.\\
2.\ Approximate test: $X_1, \ldots, X_n \sim B(1, \pi)$ i.i.d., $n \geq 30$}

\mytab{Hypothesis}{$H_0\colon~ \pi = 0.5$}

\begin{lstlisting}
## 1.
n <- 9000                                 # sample size
pval <- replicate(5000, {                 # replications of experiment
  x <- rbinom(1, size = n,                # data-generating model with
              prob = 0.5 + 0.015)         #   minimum relevant effect
  binom.test(x, n = n, p = 0.5)$p.value   # p-value of test against H0
})
mean(pval < 0.05)                         # simulated power at alpha = 0.05

## 2.
n <- 9000
pval <- replicate(5000, {
  x <- rbinom(1, size = n, prob = 0.5 + 0.015)
  prop.test(x, n = n, p = 0.5, correct = FALSE)$p.value
})
mean(pval < 0.05)
\end{lstlisting}

\subsection{One-sample $z$  and $t$ test}\label{sec:onesamplet}

\mytab{Context}{In a listening experiment, a participant will repeatedly
adjust the frequency ($X$) of a comparison tone to sound equal in pitch to a
1000-Hz standard tone. The mean adjustment estimates the point of subjective
equality $\mu$. Goal is to detect a deviation from $\mu = 1000$\,Hz by 4\,Hz.}

\mytab{Assumptions}
{1.\ $z$ test: $X_1, \ldots, X_n \sim N(\mu, \sigma^2)$ i.i.d.,
    $\sigma^2$ known\\
2.\ $t$ test: $X_1, \ldots, X_n \sim N(\mu, \sigma^2)$ i.i.d.,
    $\sigma^2$ unknown}

\mytab{Hypothesis}{$H_0\colon~ \mu = 1000$}

\begin{lstlisting}
## 1.
n <- 30
pval <- replicate(2000, {
  x <- rnorm(n, mean = 1000 + 4, sd = 7.5)
  z <- (mean(x) - 1000)/7.5*sqrt(n)         # test statistic, variance known
  2*pnorm(-abs(z))                          # p-value
})
mean(pval < 0.05)

## 2.
n <- 30
pval <- replicate(2000, {
  x <- rnorm(n, mean = 1000 + 4, sd = 7.5)
  t.test(x, mu = 1000)$p.value              # variance unknown
})
mean(pval < 0.05)
\end{lstlisting}

\subsection{Variance test}

\mytab{Context}{See \ref{sec:onesamplet}. Goal is to detect a deviation from
$\sigma = 7.5$\,Hz by 2.5\,Hz.}

\mytab{Assumptions}{$X_1, \ldots, X_n \sim N(\mu, \sigma^2)$ i.i.d.}

\mytab{Hypothesis}{$H_0\colon~ \sigma^2 = 7.5^2$}

\begin{lstlisting}
n <- 50
pval <- replicate(2000, {
  x <- rnorm(n, mean = 1000, sd = 7.5 + 2.5)
  t <- (n - 1)*var(x)/7.5^2
  p <- pchisq(t, n - 1)
  2*min(p, 1 - p)                     # p-value, two-sided
# t < qchisq(.025, n - 1) || t > qchisq(.975, n - 1)
})
mean(pval < 0.05)
\end{lstlisting}

\subsection{Sign test and Wilcoxon signed-rank test}

\mytab{Context}{In a student population, an intelligence test will be
conducted and the test score ($X$) be observed. Goal is to detect a deviation
from the median $\tilde{\mu} = 100$ by 30 or 25 points.}

\mytab{Assumptions}{$X_1, \ldots, X_n$ i.i.d.\\
1.\ Sign test: continuous distribution\\
2.\ Wilcoxon signed-rank test: continuous, symmetric distribution}

\mytab{Hypothesis}{$H_0\colon~ \tilde{\mu} = 100$}

\begin{lstlisting}
## 1.
sign.test <- function(x, median) {
  n <- length(x)
  smaller <- sum(x < median)
  equal <- sum(x == median)
  count <- min(smaller, n - equal - smaller)
  2*pbinom(count, n - equal, 0.5)
}
n <- 75
pval <- replicate(2000, {
  x <- rlnorm(n, meanlog = log(100 + 30),  # log-normal distribution, median = 130
              sdlog = 0.6)
  sign.test(x, median = 100)
})                                            
mean(pval < 0.05)

## 2.
n <- 32
pval <- replicate(2000, {
  x <- runif(n, min = 50, max = 200)       # uniform distribution, E(X) = 125
  wilcox.test(x, mu = 100)$p.value
})
mean(pval < 0.05)
\end{lstlisting}

\subsection{$\chi^2$ goodness-of-fit test}

\mytab{Context}{First example: A presumably loaded die will repeatedly be
thrown and the number of times ($X$) each side comes up will be observed. Goal
is to detect a deviation from $p_k = 1/6$ of $p_6 = 1/4$ of showing $\{6\}$
(and consequently lower probabilities of showing $\{1\}$, $\{2\}$, $\{3\}$,
$\{4\}$, or $\{5\}$).\\
Other examples: An intelligence test score ($X$) will be observed. Goal is to
detect a deviation from a specific $N(\mu = 100, \sigma^2 = 15^2)$ or from an
unspecific normal distribution by some minimum relevant amount.}

\mytab{Assumptions}{$X_1, \ldots, X_n$ arbitrary i.i.d.\ with domain $\{1,
\ldots, r\}$; continuous variables $X$ have to be categorized first}

\mytab{Hypothesis}{$H_0\colon~ P(X_i = k) = p_k$ for all $k = 1, \ldots, r$
($i = 1, \ldots, n$)}

\begin{lstlisting}
## Goodness-of-fit test of multinomial distribution
n <- 300
pval <- replicate(2000, {
  x <- rmultinom(1, size = n, prob = c(rep(0.15, 5), 0.25))
  chisq.test(x, p = rep(1/6, 6))$p.value              # H0: pk = 1/6 for all k
})
mean(pval < 0.05)

## Goodness-of-fit test of normal distribution
n <- 140
pval <- replicate(2000, {
  x <- rnorm(n, mean = 100, sd = 20)
  y <- table(cut(x, breaks = c(-Inf, seq(80, 120, 5), Inf)))
  p <- pnorm(c(seq(80, 120, 5), Inf),
             mean = 100, sd = 15)                     # parameters known
  chisq.test(y, p = c(p[1], diff(p)))$p.value
})
mean(pval < 0.05)

## Goodness-of-fit test of log-normal distribution
n <- 850
pval <- replicate(2000, {
  x <- rlnorm(n, meanlog = log(100) - 0.25^2/2, sdlog = 0.25)
  y <- table(cut(x, breaks = c(-Inf, seq(80, 120, 5), Inf)))
  p <- pnorm(c(seq(80, 120, 5), Inf),
             mean = mean(x), sd = sd(x))              # parameters unknown
  chi <- chisq.test(y, p = c(p[1], diff(p)))$statistic
  pchisq(chi, df = length(y) - 1 - 2,                 # correcting df's
         lower.tail = FALSE)
})
mean(pval < 0.05)

## Kolmogorov-Smirnov goodness-of-fit test of normal distribution
n <- 250
pval <- replicate(2000, {
  x <- rnorm(n, mean = 100, sd = 20)
  ks.test(x, "pnorm", mean = 100, sd = 15)$p.value    # parameters known
})
mean(pval < 0.05)
\end{lstlisting}

\section{Two-sample tests}

\subsection{Independent samples}

\subsubsection{Two-sample $z$, $t$ and Welch test}\label{sec:twosamplet}

\mytab{Context}{Two participants will take part in the listening experiment
from \ref{sec:onesamplet}, providing adjustments $X$ and $Y$. Goal is to
detect a difference between their points of subjective equality $\mu_x$ and
$\mu_y$ of 4\,Hz.}

\mytab{Assumptions}
{$X_1, \ldots, X_n \sim N(\mu_x, \sigma_x^2)$ i.i.d.,
$Y_1, \ldots, Y_m \sim N(\mu_y, \sigma_y^2)$ i.i.d.,
both samples independent\\
1.\ $z$ test: $\sigma_x^2$, $\sigma_y^2$ known\\
2.\ $t$ test: $\sigma_x^2 = \sigma_y^2$ but unknown\\
3.\ Welch test: $\sigma_x^2 \neq \sigma_y^2$ and unknown}

\mytab{Hypothesis}{$H_0\colon~ \mu_x - \mu_y = \delta = 0$}

\begin{lstlisting}
## 1.
n <- 85; m <- 70
pval <- replicate(2000, {
  x <- rnorm(n, mean = 1000 + 4, sd = 7)
  y <- rnorm(m, mean = 1000,     sd = 10)
  z <- (mean(x) - mean(y) - 0)/sqrt(7^2/n + 10^2/m)  # variances known
  2*pnorm(-abs(z))
})
mean(pval < 0.05)

## 2. (var.equal = TRUE), 3. (var.equal = FALSE)
n <- 115; m <- 90
pval <- replicate(2000, {
  x <- rnorm(n, mean = 1000 + 4, sd = 10)
  y <- rnorm(m, mean = 1000,     sd = 10)
  t.test(x, y, mu = 0, var.equal = TRUE)$p.value    # variances unknown
})
mean(pval < 0.05)
\end{lstlisting}

\subsubsection{Comparing two variances}

\mytab{Context}{See \ref{sec:twosamplet}. Goal is to detect that $\sigma_x^2$
is 2.5 times larger than $\sigma_y^2$.}

\mytab{Assumptions}
{$X_1, \ldots, X_n \sim N(\mu_x, \sigma_x^2)$ i.i.d.,
$Y_1, \ldots, Y_m \sim N(\mu_y, \sigma_y^2)$ i.i.d.,
both samples independent}

\mytab{Hypothesis}{$H_0\colon~ \sigma_x^2/\sigma_y^2 = 1$}

\begin{lstlisting}
n <- 45; m <- 40
pval <- replicate(2000, {
  x <- rnorm(n, mean = 1000, sd = sqrt(7^2 * 2.5))
  y <- rnorm(m, mean = 1000, sd = sqrt(7^2))
  var.test(x, y, ratio = 1)$p.value
})
mean(pval < 0.05)
\end{lstlisting}

\subsubsection{Wilcoxon rank-sum (Mann-Whitney) test}

\mytab{Context}{In two different student populations, for example, science and
humanities, an intelligence test will be conducted and the test scores $X$ and
$Y$ be observed. Goal is to detect a difference in median scores
$\tilde{\mu}_x$ and $\tilde{\mu}_y$ of 40 points.}

\mytab{Assumptions}
{$X_1, \ldots, X_n$ and $Y_1, \ldots, Y_m$ continuous i.i.d., both samples
independent}

\mytab{Hypothesis}{$H_0\colon~ \tilde{\mu}_x = \tilde{\mu}_y$}

\begin{lstlisting}
n <- 60; m <- 50
pval <- replicate(2000, {
  x <- rlnorm(n, meanlog = log(100 + 40), sdlog = 0.6)  # log-normal distribution
  y <- rlnorm(m, meanlog = log(100),      sdlog = 0.6)
  wilcox.test(x, y, mu=0)$p.value
})
mean(pval < 0.05)
\end{lstlisting}

\subsubsection{Randomization test}

\mytab{Context}{See~\ref{sec:twosamplet}}

\mytab{Assumptions}
{$X_1, \ldots, X_n$ and $Y_1, \ldots, Y_m$ continuous i.i.d., both samples
independent}

\mytab{Hypothesis}{$H_0\colon$ No effect over and above random assignment}

\begin{lstlisting}
n <- 40; m <- 35
grp <- rep(c("x", "y"), c(n, m))
pval <- replicate(500, {
  x <- c(rnorm(n, mean = 1000 + 4, sd = 5),
         rnorm(m, mean = 1000,     sd = 5))
  p0 <- t.test(x ~ grp, mu = 0)$p.value
  ps <- replicate(800,
                  t.test(sample(x) ~ grp, mu = 0)$p.value)
  mean(ps <= p0)                            # p-value of randomization test
})
mean(pval < 0.05)
\end{lstlisting}

\subsection{Dependent samples}

\subsubsection{Paired two-sample $t$ test}\label{sec:pairtwosamplet}

\mytab{Context}{Each participant will be tested twice using two slightly
different versions of an intelligence test, and the test scores $X$ and $Y$
will be observed. Goal is to detect a difference in mean scores $\mu_x$ and
$\mu_y$ of 5 points.}

\mytab{Assumptions}
{$X_1, \ldots, X_n$ and $Y_1, \ldots, Y_n$ samples where $D_1 = X_1 - Y_1,
\ldots, D_n = X_n - Y_n \sim N(\mu_D, \sigma_D^2)$, $\sigma_D^2 = \sigma_x^2 + \sigma_y^2 - 2\sigma_{xy}$ unknown}

\mytab{Hypothesis}{$H_0\colon~ \mu_D = 0$}

\begin{lstlisting}
n <- 18                                            # number of pairs
s <- 15; r <- 0.9; sxy <- r*s*s                    # sd of X and Y; cor; cov

## 1. Bivariate normal distribution
pval <- replicate(2000, {
  x <- MASS::mvrnorm(n, mu = c(100 + 5, 100),
                     Sigma = matrix(c(s^2, sxy, sxy, s^2), 2, 2))
  t.test(x[, 1], x[, 2], mu = 0, paired = TRUE)$p.value
})
mean(pval < 0.05)

## 2. Normal distribution of differences
pval <- replicate(2000, {
  d <- rnorm(n, mean = 5, sd = sqrt(s^2 + s^2 - 2*sxy))
  t.test(d, mu = 0)$p.value
})
mean(pval < 0.05)
\end{lstlisting}

\subsubsection{Paired two-sample Wilcoxon test}

\mytab{Context}{See \ref{sec:pairtwosamplet}.}

\mytab{Assumptions}
{$X_1, \ldots, X_n$, $Y_1, \ldots, Y_n$ samples where $D_1 = X_1 - Y_1,
\ldots, D_n = X_n - Y_n$ continuous, symmetric i.i.d.}

\mytab{Hypothesis}{$H_0\colon~ \tilde{\mu}_D = 0$}

\begin{lstlisting}
n <- 18
s <- 15; r <- 0.9; sxy <- r*s*s
pval <- replicate(2000, {
  x <- MASS::mvrnorm(n, mu = c(100 + 5, 100),
                     Sigma = matrix(c(s^2, sxy, sxy, s^2), 2, 2))
  wilcox.test(x[, 1], x[, 2], mu = 0, paired = TRUE)$p.value
})
mean(pval < 0.05)
\end{lstlisting}

\subsubsection{Paired two-sample randomization test}

\mytab{Context}{See \ref{sec:pairtwosamplet}.}

\mytab{Assumptions}
{$X_1, \ldots, X_n$, $Y_1, \ldots, Y_n$ samples where $D_1 = X_1 - Y_1,
\ldots, D_n = X_n - Y_n$ continuous i.i.d.}

\mytab{Hypothesis}{$H_0\colon$ No effect over and above randomization of order}

\begin{lstlisting}
n <- 18                                            # number of pairs
s <- 15; r <- 0.9; sxy <- r*s*s                    # sd of X and Y; cor; cov
pval <- replicate(500, {
  d <- rnorm(n, mean = 5, sd = sqrt(s^2 + s^2 - 2*sxy))
  p0 <- t.test(d, mu = 0)$p.value
  ps <- replicate(800,
                  t.test(sample(c(-1, 1), n, replace = TRUE) * (d - 0))$p.value)
  mean(ps <= p0)    # p-value of randomization test
})
mean(pval < 0.05)
\end{lstlisting}

\section{Analysis of association}

\subsection{$\chi^2$ test of homogeneity}

\mytab{Context}{Two courses will be compared regarding the grades (1 to 5)
they produce ($X$). Goal is to detect a deviation from homogeneity where the
first course produces less extreme grades than the second course.}

\mytab{Assumptions}
{$X_{j1}, \ldots, X_{j n_j}$ independent and identically multinomially
distributed (domain $\{1, \ldots, r\}$) for all independent samples $j = 1,
\ldots, m$}

\mytab{Hypothesis}{$H_0\colon~ 
P(X_{ji} = k) = p_k$ for all $j=1, \ldots, m$, $i = 1, \ldots,
                                           n_j$ und $k = 1, \ldots, r$}

\begin{lstlisting}
n1 <- 250; n2 <- 190
pval <- replicate(2000, {
  x <- cbind(                           # conditional probabilities
         rmultinom(1, size = n1, prob = c(.09, .25, .32, .25, .09)),
         rmultinom(1, size = n2, prob = c(.16, .22, .24, .22, .16))
  )
  chisq.test(x)$p.value
})
mean(pval < 0.05)
\end{lstlisting}

\subsection{$\chi^2$ test of independence}

\mytab{Context}{A single population will be cross-classified by two
categorical variables $X$ and $Y$ having a specific joint distribution.
Alternatively, two correlated continuous variables are assumed to underlie $X$
and $Y$. Goal is to detect certain combinations that are more or less frequent
than expected under independence of $X$ and $Y$.}

\mytab{Assumptions}
{Independent pairs $(X_i,Y_i)$, $i = 1, \ldots, n$, where $X_i \in \{1,
\ldots, m\}$ and $Y_i \in \{1, \ldots, r\}$}

\mytab{Hypothesis}{$H_0\colon~ 
P(X_i = j, Y_i = k) = P(X_i = j) \cdot P(Y_i = k)$ for all
                                           $j,k$ and $i = 1, \ldots, n$}

\begin{lstlisting}
## 1. Joint multinomial distribution
n <- 90                                                # number of pairs
pval <- replicate(2000, {
  dat <- cbind(expand.grid(1:2, 1:3), x = rmultinom(1, size = n,
           prob = c(.10, .23, .17, .17, .23, .10)))    # joint probabilities
  tab <- xtabs(x ~ ., data = dat)
  chisq.test(tab)$p.value
})
mean(pval < 0.05)

## 2. Latent continuous variables
n <- 130
pval <- replicate(2000, {
  x <- MASS::mvrnorm(n, mu = c(0, 0),
                     Sigma = matrix(c(1, 0.4, 0.4, 1), 2, 2))
  tab <- table(cut(x[, 1], breaks = 2), cut(x[, 2], breaks = 3))
  chisq.test(tab)$p.value
})
mean(pval < 0.05)
\end{lstlisting}

\subsection{Test of correlation}

\mytab{Context}{Each participant will be tested twice using two intelligence
tests, and the test scores $X$ and $Y$ will be observed and their correlation
$\rho$ be estimated. Goal is to detect a deviation from $\rho = 0$ or from
$\rho = 0.6$ by 0.3.}

\mytab{Assumptions}
{$(X_i, Y_i)$ independent jointly normally distributed, $i = 1, \ldots, n,$
\[
  \hat{\rho}_{XY} = r_{XY} = \frac{\sum_{i=1}^n (X_i - \bar{X})(Y_i - \bar{Y})}
  {\sqrt{\sum_{i=1}^n (X_i - \bar{X})^2 \cdot \sum_{i=1}^n (Y_i - \bar{Y})^2}}
\]}
\mytab{Hypothesis}{$H_0\colon~ \rho_{XY} = \rho_0$}

\begin{lstlisting}
## rho0 = 0
n <- 90
s <- 15; r <- 0.3; sxy <- r*s*s
pval <- replicate(2000, {
  x <- MASS::mvrnorm(n, mu = c(100, 100),
                     Sigma = matrix(c(s^2, sxy, sxy, s^2), 2, 2))
  cor.test(x[, 1], x[, 2])$p.value
})
mean(pval < 0.05)

## rho0 = 0.6
n <- 60
s <- 15; r <- 0.3; sxy <- r*s*s
rho0 <- 0.6
pval <- replicate(2000, {
  x <- MASS::mvrnorm(n, mu = c(100, 100),
                     Sigma = matrix(c(s^2, sxy, sxy, s^2), 2, 2))
  z <- 1/2 * (log((1 + cor(x[, 1], x[, 2])) / (1 - cor(x[, 1], x[, 2]))) -
              log((1 + rho0) / (1 - rho0))) * sqrt(n - 3)
  2*pnorm(-abs(z))
})
mean(pval < 0.05)
\end{lstlisting}

\section{Regression}

\subsection{Simple regression}\label{sec:simplereg}

\mytab{Context}{In a group of patients, the strength of depressive symptoms
($Y$) will be measured. Each patient will receive 0, 2, \dots, 8\,mg dose
($X$) of medication. Goal is to detect a linear reduction of depressive
symptoms of 2.5 points per mg dose.}

\mytab{Model}
{$
  Y_i = \alpha + \beta X_i + \varepsilon_i, \quad
  \varepsilon_i \sim N(0, \sigma^2) \text{ i.i.d.}, \quad i = 1, \dots, n
$}
\mytab{Hypothesis}{$H_0\colon~ \beta = 0$}

\begin{lstlisting}
## Wald test
n <- 35
x <- rep(c(0, 2, 4, 6, 8), each = n/5)
pval <- replicate(2000, {
  y <- 35 - 2.5*x + rnorm(n, sd = 14)  # y = a + b*x + e
  m <- lm(y ~ x)
  coef(summary(m))["x", "Pr(>|t|)"]
})
mean(pval < 0.05)

## Likelihood ratio (overall F) test
n <- 35
x <- rep(c(0, 2, 4, 6, 8), each = n/5)
pval <- replicate(2000, {
  y <- 35 - 2.5*x + rnorm(n, sd = 14)
  m1 <- lm(y ~ 1)                      # null model
  m2 <- lm(y ~ x)
  anova(m1, m2)$"Pr(>F)"[2]
})
mean(pval < 0.05)
\end{lstlisting}

\subsection{Multiple regression}\label{sec:multireg}

\mytab{Context}{A similar study as in \ref{sec:simplereg} will be conducted,
but participants will be sampled from three different age groups $X_2$. Goal
is to detect either a linear reduction of symptoms of 2.5 points per mg dose
in each age group, or a linear increase of 0.1 points per year of age in each
medication group, or both.}

\mytab{Model}
{$
  Y_i = \beta_0 + \beta_1 \cdot X_{i1} + \ldots + \beta_p \cdot X_{ip} +
        \varepsilon_i, \quad
  \varepsilon_i \sim N(0, \sigma^2) \text{ i.i.d.}, \quad i = 1, \dots, n
$}
\mytab{Hypothesis}{$H_0\colon~ \beta_1 = \beta_2 = 0$}

\begin{lstlisting}
n <- 30
dat <- expand.grid(x1 = rep(c(0, 2, 4, 6, 8), each = n/(5*3)),
                   x2 = c(30, 50, 80))
pval <- replicate(2000, {
  y <- 35 - 2.5*dat$x1 + 0.1*dat$x2 + rnorm(n, sd = 14)
  m1 <- lm(y ~ 1,       dat)
  m2 <- lm(y ~ x1 + x2, dat)
  anova(m1, m2)$"Pr(>F)"[2]
})
mean(pval < 0.05)

## Parameter recovery
n <- 3000
dat <- expand.grid(x1 = rep(c(0, 2, 4, 6, 8), each = n/(5*3)),
                   x2 = c(30, 50, 80))
y <- 35 - 2.5*dat$x1 + 0.1*dat$x2 + rnorm(n, sd = 14)
m <- lm(y ~ x1 + x2, dat)
c(coef(m), sigma(m))
lattice::xyplot(y ~ x1, dat, groups = x2, type = c("g", "p", "r"))
\end{lstlisting}

\subsection{Binomial regression}

\mytab{Context}{On each trial of a psychophysics experiment, a participant
will see a 40-cd/m$^2$ standard light and a comparison with intensities ($X$)
ranging from 37 to 43\,cd/m$^2$. His task is to judge if the comparison is
brighter than the standard. Goal is to detect an increase of the odds of a yes
response ($Y = 1$) by a factor of 1.5 per cd/m$^2$ of comparison intensity.}

\mytab{Model}
{$
  \log \dfrac{P(Y_i = 1)}{1 - P(Y_i = 1)} = \alpha + \beta X_i, \quad
  Y_i \sim B(1, \pi_i) \text{ i.i.d.}, \quad i = 1, \dots, n
$}
\mytab{Hypothesis}{$H_0\colon~ \beta = 0$}

\begin{lstlisting}
n <- 70
x <- 37:43
logit <- -40*log(1.5) + x*log(1.5)                 # odds ratio = 1.5
pval <- replicate(2000, {
  y <- rbinom(7, size = n/7, prob = plogis(logit))
  m1 <- glm(cbind(y, n/7 - y) ~ 1, binomial)
  m2 <- glm(cbind(y, n/7 - y) ~ x, binomial)
  anova(m1, m2, test = "LRT")$"Pr(>Chi)"[2]
})
mean(pval < 0.05)
\end{lstlisting}

\section{Analysis of variance}

\subsection{One-way analysis of variance}

\subsubsection{Model with fixed effects}

\mytab{Context}{Two alternative treatments ($i = 2$, $i = 3$) of a disease
will be compared to a control group ($i = 1$) and the symptoms $Y$ be
measured. Goal is to detect a deviation from the control group of 2 for
Treatment~1 and of $-3$ for Treatment~2.}

\mytab{Model}
{$
  Y_{ij} = \mu + \alpha_i + \varepsilon_{ij}, \quad
  \varepsilon_{ij} \sim N(0, \sigma^2) \text{ i.i.d.}
$,
$i = 1, \dots, m$ (factor levels),
$j = 1, \ldots, n_i$ (observations per level),
$\alpha_1 := 0$}

\mytab{Hypothesis}{$H_0\colon~ \alpha_2 = \alpha_3 = 0$}

\begin{lstlisting}
n1 <- 22; n2 <- 24; n3 <- 18
grp <- factor(rep(1:3, c(n1, n2, n3)), labels = c("ctl", "t1", "t2"))
beta <- c(mu = 10, a2 = 2, a3 = -3)                          # a1 := 0
means <- model.matrix(~ grp) %*% beta
pval <- replicate(2000, {
  y <- means + rnorm(n1 + n2 + n3, sd = 5)                   # y = mu + a + e
  m <- aov(y ~ grp)
  summary(m)[[1]]$"Pr(>F)"[1]
})
mean(pval < 0.05)
\end{lstlisting}

\mytab{Context}{In a similar study, it is the goal to detect a deviation from
the control group of the treatment average, assuming an effect of 3 for
Treatment~1 and of $4.5$ for Treatment~2.}

\mytab{Assumptions}
{Linear contrasts:
$
  \sum_{i = 1}^m c_i \cdot \bar{Y}_i, \text{ where } \sum_i c_i = 0
$\\
Orthogonality: $\sum_i c_{ij} \cdot c_{ik} = 0$ for pairs $(j, k)$ of
contrasts}

\mytab{Hypothesis}{$H_0\colon~ -1\cdot \mu_1 + \tfrac{1}{2}\cdot \mu_2 +
                       \tfrac{1}{2}\cdot \mu_3 = 0$}

\begin{lstlisting}
n <- 72
grp <- factor(rep(1:3, each = n/3), labels = c("ctl", "t1", "t2"))
beta <- c(mu = 10, a2 = 3, a3 = 4.5)                         # a1 := 0
means <- model.matrix(~ grp) %*% beta
contrasts(grp) <- cbind(c(-1, 1/2, 1/2),
                        c( 0,  -1,   1))
pval <- replicate(2000, {
  y <- means + rnorm(n, sd = 5)                              # y = mu + a + e
  m <- aov(y ~ grp)
  summary(m,
          split = list(grp = list("c-1,2" = 1, "1-2" = 2)))[[1]]$"Pr(>F)"["c-1,2"]
})
mean(pval < 0.05)
\end{lstlisting}

\subsubsection{Model with random effects}

\mytab{Context}{In an experiment, each participant will react to one of ten
different items ($A$) by pressing a button. Reaction time in ms ($Y$) will be
measured. Goal is to detect an amount of variation between items of $\sigma_A
= 10$\,ms.}

\mytab{Model}
{$
  Y_{ij} = \mu + \alpha_i + \varepsilon_{ij}
$, \quad
$\alpha_i \sim N(0, \sigma_A^2)$,
$\varepsilon_{ij} \sim N(0, \sigma^2)$,
all random variables independent}

\mytab{Hypothesis}{$H_0\colon~ \sigma_A^2 = 0$}

\begin{lstlisting}
n <- 60
item <- factor(rep(1:10, n/10))
pval <- replicate(2000, {
  y <- 120 + model.matrix(~ 0 + item) %*% rnorm(10, sd = 10) +
             rnorm(n, sd = 15)                               # y = mu + a + e
  m <- aov(y ~ Error(item))
  ms <- sapply(summary(m), function(x) x[[1]]$"Mean Sq")
  df <- sapply(summary(m), function(x) x[[1]]$"Df")
  pf(ms["Error: item"]/ms["Error: Within"],
     df["Error: item"], df["Error: Within"], lower.tail = FALSE)
})
mean(pval < 0.05)

## Parameter recovery
n <- 4000
item <- factor(rep(1:10, n/10))
y <- 120 + model.matrix(~ 0 + item) %*% rnorm(10, sd = 10) +
           rnorm(n, sd = 15)
m <- aov(y ~ Error(item))
ms <- sapply(summary(m), function(x) x[[1]]$"Mean Sq")
c(
  mu = coef(m)$"(Intercept)",
  sa = sqrt((ms["Error: item"] - ms["Error: Within"])/(n/10)),
   s = sqrt(ms["Error: Within"])
)
\end{lstlisting}

\subsection{Two-way analysis of variance}

\subsubsection{Model with fixed effects}

\mytab{Context}{In an experiment, two fertilizers ($A$ and $B$, each either
low or high dose) will be combined and the yield of peas ($Y$) in kg be
observed. Goal is to detect an increase of the Fertilizer-A effect by an
additional 12\,kg when combined with a high dose of Fertilizer B (interaction
effect).}

\mytab{Model}
{$
  Y_{ijk} = \mu + \alpha_i + \beta_j + (\alpha\beta)_{ij} + \varepsilon_{ijk},
  \quad \varepsilon_{ijk} \sim N(0, \sigma^2) \text{ i.i.d.}
$,\\
$i = 1, \dots, I$; $j = 1, \dots, J$; $k = 1, \dots, K$;
$\alpha_1 = \beta_1 := 0$}

\mytab{Hypothesis}
{$H_0^{AB}\colon~ (\alpha\beta)_{ij} = 0 \text{ for all } i,j$}

\begin{lstlisting}
n <- 96
dat <- data.frame(
  A = factor(rep(1:2, each = n/2), labels = c("low", "high")),
  B = factor(rep(rep(1:2, each = n/4), 2), labels = c("low", "high"))
)
beta <- c(mu = 30, a2 = 30, b2 = 5, ab22 = 12)
means <- model.matrix(~ A*B, dat) %*% beta
pval <- replicate(2000, {
  y <- means + rnorm(n, sd = 10)                     # y = mu + a + b + ab + e
  m <- aov(y ~ A*B, dat)
  summary(m)[[1]]$"Pr(>F)"[3]                        # test of interaction
})
mean(pval < 0.05)
\end{lstlisting}

\subsubsection{Model with random effects}

\mytab{Context}{Eight test administrators ($B$) conduct the Rorschach test in
six different orders ($A$) with a new sample of patients for each combination.
Goal is to detect if certain combinations lead to higher test scores, which is
reflected by a variation of the interaction effect of $\sigma_{AB} = 7$
points.}

\mytab{Model}
{$
  Y_{ijk} = \mu + \alpha_i + \beta_j + (\alpha\beta)_{ij} + \varepsilon_{ijk}
$, \quad
$\alpha_i \sim N(0, \sigma_A^2)$,
$\beta_j \sim N(0, \sigma_B^2)$,
$(\alpha\beta)_{ij} \sim N(0, \sigma_{AB}^2)$,
$\varepsilon_{ijk} \sim N(0, \sigma^2)$,
all random variables independent}

\mytab{Hypothesis}{$H_0^{AB}\colon~ \sigma_{AB}^2 = 0$}

\begin{lstlisting}
n <- 192
dat <- expand.grid(order = factor(rep(1:6, each = n/(6*8))),
                   admin = factor(1:8))
Z <- model.matrix(~ 0 + order*admin, dat,
                  contrasts.arg = lapply(dat, contrasts, contrasts = FALSE))
pval <- replicate(2000, {
  y <- 50 + Z %*% c(rnorm(6, sd = 10), rnorm(8, sd = 15), rnorm(6*8, sd = 7)) +
            rnorm(n, sd = 13)                         # y = mu + a + b + ab + e
  m <- aov(y ~ Error(order*admin), dat)
  ms <- sapply(summary(m), function(x) x[[1]]$"Mean Sq")
  df <- sapply(summary(m), function(x) x[[1]]$"Df")
  pf(ms["Error: order:admin"]/ms["Error: Within"],
     df["Error: order:admin"], df["Error: Within"], lower.tail = FALSE)
})
mean(pval < 0.05)

## Parameter recovery
n <- 9600; k <- n/(6*8)
dat <- expand.grid(order = factor(rep(1:6, each = k)),
                   admin = factor(1:8))
Z <- model.matrix(~ 0 + order*admin, dat,
                  contrasts.arg = lapply(dat, contrasts, contrasts = FALSE))
y <- 50 + Z %*% c(rnorm(6, sd = 10), rnorm(8, sd = 15), rnorm(6*8, sd = 7)) +
          rnorm(n, sd = 13)
m <- aov(y ~ Error(order*admin), dat)
ms <- sapply(summary(m), function(x) x[[1]]$"Mean Sq")
c(
   mu = coef(m)$"(Intercept)",
   sa = sqrt((ms["Error: order"] - ms["Error: order:admin"])/(8*k)),
   sb = sqrt((ms["Error: admin"] - ms["Error: order:admin"])/(6*k)),
  sab = sqrt((ms["Error: order:admin"] - ms["Error: Within"])/k),
    s = sqrt(ms["Error: Within"])
)
\end{lstlisting}

\subsection{Repeated measures ANOVA}

\subsubsection{Model with one repeated-measures factor}

\mytab{Context}{In an experiment, each participant ($P$) will complete a
mirror-drawing task with four shapes: circle, triangle, square, and star
($B$). The number of errors will be measured. Goal is to detect a difference
to the circle of 2, 4, and 6 errors.}

\mytab{Model}
{$
  Y_{ij} = \mu + \pi_i + \beta_j + (\pi\beta)_{ij}, \quad
  i = 1, \dots, I \text{ (number of subjects)}, j = 1, \dots, J
$,
$\pi_i \sim N(0, \sigma_P^2)$,
$(\pi\beta)_{ij} \sim N(0, \sigma_{PB}^2)$,
all random variables independent,
$\beta_1 := 0$}

\mytab{Hypothesis}{$\text{H$_0$: } \beta_2 = \beta_3 = \beta_4 = 0$}

\begin{lstlisting}
n <- 22                                             # number of subjects
dat <- expand.grid(shape = factor(rep(1:4),
                                  labels = c("circle", "triangle",
                                             "square", "star")),
                    subj = factor(1:n))
beta <- c(mu = 15, b2 = 2, b3 = 4, b4 = 6)          # b1 := 0
means <- model.matrix(~ shape, dat) %*% beta
pval <- replicate(2000, {
  y <- means +
       model.matrix(~ 0 + subj, dat) %*% rnorm(n, sd = 4) +
       rnorm(4*n, sd = 6)                           # y = mu + p + b + e
  m <- aov(y ~ shape + Error(subj), dat)
  summary(m)$"Error: Within"[[1]]$"Pr(>F)"[1]
})
mean(pval < 0.05)

## Parameter recovery
n <- 900
dat <- expand.grid(shape = factor(rep(1:4),
                                  labels = c("circle", "triangle",
                                             "square", "star")),
                    subj = factor(1:n))
beta <- c(mu = 15, b2 = 2, b3 = 4, b4 = 6)
means <- model.matrix(~ shape, dat) %*% beta
y <- means +
     model.matrix(~ 0 + subj, dat) %*% rnorm(n, sd = 4) +
     rnorm(4*n, sd = 6)
m <- aov(y ~ shape + Error(subj), dat)
ms <- sapply(summary(m), function(x) x[[1]]["Residuals", "Mean Sq"])
c(
   b = coef(m)$"Within",
  sp = sqrt((ms["Error: subj"] - ms["Error: Within"])/4),
   s = sqrt(ms["Error: Within"])
)
\end{lstlisting}

\subsubsection{Model with two repeated-measures factors}

\mytab{Context}{Participants ($P$) will be presented with positive and
negative sentences ($A$) that contain the pronoun ``I'' or ``You'' or
``He/She'' ($B$). Each participant will read all six combinations, and the
reading time will be measured. Goal is to detect an attenuation of the
negation effect of 150\,ms for ``You'' and ``He/She'' sentences as compared to
``I'' sentences (interaction effect).}

\mytab{Model}
{$
  Y_{ij} = \mu + \alpha_j + \beta_k + (\alpha\beta)_{jk} +
  \pi_i + (\pi\alpha)_{ij} + (\pi\beta)_{ik} + (\pi\alpha\beta)_{ijk}, \quad
  i = 1, \dots, I;~ j = 1, \dots, J;~ k = 1, \dots, K
$,
$\pi_i                  \sim N(0, \sigma_P^2)$,
$(\pi\alpha)_{ij}       \sim N(0, \sigma_{PA}^2)$,
$(\pi\beta)_{ik}        \sim N(0, \sigma_{PB}^2)$,
$(\pi\alpha\beta)_{ijk} \sim N(0, \sigma^2)$,
all random variables independent,
$\alpha_1 = \beta_1 := 0$}

\mytab{Hypothesis}{$H_0^{AB}\colon~ (\alpha\beta)_{jk} = 0 \text{ for all }
j,k$}

\begin{lstlisting}
n <- 136
dat <- expand.grid(A = factor(1:2), B = factor(1:3), subj = factor(1:n))
beta <- c(mu = 1500, a2 = 300, b2 = 200, b3 = 250, ab22 = -150, ab23 = -150)  # a1, b1 := 0
X <- model.matrix(~ A*B, dat)
Z <- model.matrix(~ 0 + subj + subj:A + subj:B, dat,
                  contrasts.arg = lapply(dat, contrasts, contrasts = FALSE))
pval <- replicate(2000, {
  y <- X %*% beta +                 # y = mu + a + b + ab + p + pa + pb + e
       Z %*% c(p = rnorm(n, sd = 350), pa = rnorm(2*n, sd = 120),
               pb = rnorm(3*n, sd = 80)) +
       rnorm(6*n, sd = 300)
  m <- aov(y ~ A*B + Error(subj + subj:A + subj:B), dat)
  summary(m)$"Error: Within"[[1]]$"Pr(>F)"[1]
})
mean(pval < 0.05)

## Parameter recovery
n <- 620
dat <- expand.grid(A = factor(1:2), B = factor(1:3), subj = factor(1:n))
beta <- c(mu = 1500, a2 = 300, b2 = 200, b3 = 250, ab22 = -150, ab23 = -150)
X <- model.matrix(~ A*B, dat)
Z <- model.matrix(~ 0 + subj + subj:A + subj:B, dat,
                  contrasts.arg = lapply(dat, contrasts, contrasts = FALSE))
y <- X %*% beta +
     Z %*% c(p = rnorm(n, sd = 350), pa = rnorm(2*n, sd = 120),
             pb = rnorm(3*n, sd = 80)) +
     rnorm(6*n, sd = 300)
m <- aov(y ~ A*B + Error(subj + subj:A + subj:B), dat)
ms <- sapply(summary(m), function(x) x[[1]]["Residuals", "Mean Sq"])
c(
   ab = coef(m)$"Within",
   sp = sqrt((ms["Error: subj"] - ms["Error: subj:A"] -
              ms["Error: subj:B"] + ms["Error: Within"])/6),
  spa = sqrt((ms["Error: subj:A"] - ms["Error: Within"])/3),
  spb = sqrt((ms["Error: subj:B"] - ms["Error: Within"])/2),
    s = sqrt(ms["Error: Within"])
)
\end{lstlisting}

\section{Analysis of covariance}

\mytab{Context}{High school students will be randomized to three different
programming courses ($A$) and their performance will be tested in a final exam
($Y$). Prior to randomization, their computer knowledge will be measured in a
pre-test ($X$). Goal is to detect a pre-test adjusted deviation from Course~1
of 0.5 and 4 points for Course~2 and 3.}

\mytab{Model}{
$
  Y_{ij} = \mu + \alpha_i + \beta \cdot X_{ij} + \varepsilon_{ij}, \quad
  \varepsilon_{ij} \sim N(0, \sigma^2) \text{ i.i.d.},
  i = 1, \dots, m;~ j = 1, \dots, n,
  \alpha_1 := 0
$}

\mytab{Hypothesis}{$H_0\colon~ \alpha_2 = \alpha_3 = 0$}

\begin{lstlisting}
## Likelihood ratio (incremental F) test
n <- 84
dat <- data.frame(
  pre = rnorm(n, mean = 20, sd = 8),
  grp = factor(rep(1:3, each = n/3))
)
beta <- c(mu = 5, b = 0.7, a2 = 0.5, a3 = 4)            # a1 := 0
means <- model.matrix(~ pre + grp, dat) %*% beta
pval <- replicate(2000, {
  y <- means + rnorm(n, sd = 5)                         # y = mu + a + b*x + e
  m1 <- lm(y ~ pre,       dat)
  m2 <- lm(y ~ pre + grp, dat)
  anova(m1, m2)$"Pr(>F)"[2]
})
mean(pval < 0.05)
\end{lstlisting}

\section{Multivariate tests}

\subsection{Multivariate one-sample $t$ test}

\mytab{Context}{Participants will perform four intelligence subtests
($\vect{x}$ consisting of: information, similarities, arithmetic, picture
completion). Goal is to detect a deviation from $\gvect{\mu} = (12, 10, 10,
8)'$ by $(0.5, -1, 1, 0)'$ score points.}

\mytab{Assumptions}
{$\vect{x}_1, \ldots, \vect{x}_n \sim N(\gvect{\mu}, \gmat{\Sigma})$ i.i.d.}

\mytab{Hypothesis}{$H_0\colon~ \gvect{\mu} = (12, 10, 10, 8)'$}

\begin{lstlisting}
n <- 30                                        # sample size
s <- c(3.5, 3.5, 3.5, 2)                       # standard deviations

R <- diag(4)                                   # correlation matrix
R[lower.tri(R)] <- c(7, 5, 3, 5, 1, 3)/10
R[upper.tri(R)] <- t(R)[upper.tri(R)]

S <- R * s %o% s                               # covariance matrix

pval <- replicate(2000, {
  x <- MASS::mvrnorm(n, mu = c(12 + 0.5, 10 - 1, 10 + 1, 8), Sigma = S)
  m <- lm(x ~ offset(cbind(12, 10, 10, 8)[rep(1, n), ]))
  anova(m)$"Pr(>F)"[1]
})
mean(pval < 0.05)
\end{lstlisting}

\subsection{Multivariate two-sample $t$ test}

\mytab{Context}{Two populations of patients, for example, from psychiatry and
from oncology, will be compared regarding their age, blood pressure, and
cholesterol ($\vect{x}$ and $\vect{y}$). Goal is to detect a difference
between $\gvect{\mu}_x$ and $\gvect{\mu}_y$ of 10\,years, 5\,mmHg, and
15\,mg/dl.}

\mytab{Assumptions}
{$\vect{x}_1, \ldots, \vect{x}_n \sim N(\gvect{\mu}_x, \gmat{\Sigma})$ i.i.d.,
 $\vect{y}_1, \ldots, \vect{y}_n \sim N(\gvect{\mu}_y, \gmat{\Sigma})$ i.i.d.,
both samples independent}

\mytab{Hypothesis}{$H_0\colon~ \gvect{\mu}_x - \gvect{\mu}_y = \gvect{\delta}
= \vect{0}$}

\begin{lstlisting}
n1 <- n2 <- 50                                 # sample size per group
s <- c(15, 15, 44)
r <- 0.3                                       # constant correlation
S <- r * s %o% s; diag(S) <- s^2
dat <- data.frame(grp = rep(c("trt", "ctl"), c(n1, n2)))

pval <- replicate(2000, {
  y <- rbind(
    MASS::mvrnorm(n1, mu = c(45 + 10, 85 + 5, 200 + 15), Sigma = S),
    MASS::mvrnorm(n2, mu = c(45,      85,     200     ), Sigma = S)
  )
  anova(lm(y ~ grp, dat))$"Pr(>F)"[2]
})
mean(pval < 0.05)
\end{lstlisting}

\subsection{Paired multivariate two-sample $t$ test}

\mytab{Context}{Each participant will be tested twice, before ($\vect{x}$) and
after ($\vect{y}$) a training, on three outcome measures. Goal is to detect a
difference in mean scores $\gvect{\mu}_x$ and $\gvect{\mu}_y$ of $(0.5, -0.5,
0.7)'$ points.}
 
\mytab{Assumptions}
{$\vect{x}_1, \ldots, \vect{x}_n$ and $\vect{y}_1, \ldots, \vect{y}_n$
samples where $\vect{d}_1 = \vect{x}_1 - \vect{y}_1, \ldots, \vect{d}_n =
\vect{x}_n - \vect{y}_n \sim N(\gvect{\mu}_x - \gvect{\mu}_y,
\gmat{\Sigma}_d)$ i.i.d.}

\mytab{Hypothesis}{$H_0\colon~ \gvect{\mu}_x - \gvect{\mu}_y = \vect{0}$}

\begin{lstlisting}
n <- 25                                                  # number of pairs
s <- c(2, 2, 2)

R <- diag(2) %x% matrix(.2, 3, 3) +                      # correlation within
     (1 - diag(2)) %x% (matrix(.1, 3, 3) + diag(.6, 3))  #   and between groups
diag(R) <- 1
S <- R * rep(s, 2) %o% rep(s, 2)

C <- rbind(c(1, 0, 0, -1, 0, 0),
           c(0, 1, 0, 0, -1, 0),
           c(0, 0, 1, 0, 0, -1))
Sd <- C %*% S %*% t(C)                                   # covariance matrix of differences

pval <- replicate(2000, {
  d <- MASS::mvrnorm(n, mu=c(0 + 0.5, 0 - 0.5, 0 + 0.7), Sigma = Sd)
  m <- lm(d ~ 1)
  anova(m)$"Pr(>F)"[1]
})
mean(pval < 0.05)
\end{lstlisting}

\section{Mixed-effects models}

\mytab{Context}{A sample of public and private (\texttt{sector}) schools will
be drawn, and within each school a sample of students (\texttt{id}). All
students will complete a math test ($Y$). Socio-economic status for each
student (\texttt{cses}) and averaged for each school (\texttt{meanses}) will
be considered as predictors. Goal is to detect an increased \texttt{cses}
effect for higher-status schools by $\beta_3 = 0.7$ points per unit change,
and an attenuated \texttt{cses} effect by $\beta_5 = -0.7$ points for private
schools (interaction effects).}

\mytab{Model}{
\vspace{-5ex}
\begin{align*}
  Y_{ij} &= \beta_0 + \beta_1 \, \mathtt{cses}_{ij} + \beta_2 \, \mathtt{meanses}_i + \beta_4 \, \mathtt{sector}_i \\
         &  \quad + \beta_3 \, (\mathtt{cses}_{ij} \times \mathtt{meanses}_i) + \beta_5 \, (\mathtt{cses}_{ij} \times \mathtt{sector}_i) \\
         &  \quad  + \upsilon_{0i} + \upsilon_{1i}\, \mathtt{cses}_{ij}  + \varepsilon_{ij} \\
  \begin{pmatrix} \upsilon_{0i}\\ \upsilon_{1i} \end{pmatrix} &\sim
        N \left(\begin{pmatrix} 0\\ 0 \end{pmatrix}, \, \gmat{\Sigma}_\upsilon =
  \begin{pmatrix}
            \sigma^2_{\upsilon_0} & \sigma_{\upsilon_0 \upsilon_1} \\
            \sigma_{\upsilon_0 \upsilon_1} & \sigma^2_{\upsilon_1} \\
  \end{pmatrix} \right)
        \text{ i.i.d.}, \quad
  \gvect{\varepsilon}_i \sim N(\vect{0}, \, \sigma^2 \mat{I}_{n_i})
        \text{ i.i.d.} \\
  i &= 1, \ldots, I, \quad j = 1, \ldots n_i
\end{align*}}
\mytab{\vspace{-4ex}Hypothesis}
{\vspace{-4ex}$H_0\colon~ \beta_3 = \beta_5 = 0$}

\vspace{-2ex}
\lstset{texcl=true}
\begin{lstlisting}
library(lme4)
nschool <- 150
nstudent <- 40 * nschool
dat <- data.frame(
              id = seq_len(nstudent),
          school = seq_len(nschool),
          sector = rep(0:1, each = nschool/2),
             ses = rnorm(nstudent, 0, .8)
)
dat$meanses <- ave(dat$ses, dat$school)
dat$cses <- dat$ses - dat$meanses

beta <- c("(Intercept)" = 12, meanses = 5, cses = 2.5, sector = 1,
          "meanses:cses" = 0.7, "cses:sector" = -0.7)    # fixed effects
s <- 6.5                                                 # $\sigma$
su <- c(1.6, 0.3)
Su <- 0.4 * su %o% su; diag(Su) <- su^2                  # $\gmat{\Sigma}_\upsilon = \gmat{\Lambda_\vartheta} \gmat{\Lambda'_\vartheta} \sigma^2$
Lt <- chol(Su)/s
pars <- list(theta = t(Lt)[lower.tri(Lt, TRUE)], beta = beta, sigma = s)

pval <- replicate(200, {
  y <- simulate(~ meanses*cses + sector*cses + (1 + cses | school),
                newparams = pars, newdata = dat)$sim_1
  m1 <- lmer(y ~ meanses + cses + sector + (1 + cses | school), dat,
             REML = FALSE)
  m2 <- lmer(y ~ meanses*cses + sector*cses + (1 + cses | school), dat,
             REML = FALSE)
  anova(m1, m2)$"Pr(>Chisq)"[2]
})
mean(pval < .05)
\end{lstlisting}
\lstset{texcl=false}
\begin{lstlisting}
## Parameter recovery
nschool <- 200
nstudent <- 50 * nschool
dat <- data.frame(
              id = seq_len(nstudent),
          school = seq_len(nschool),
          sector = rep(0:1, each = nschool/2),
             ses = rnorm(nstudent, 0, .8)
)
dat$meanses <- ave(dat$ses, dat$school)
dat$cses <- dat$ses - dat$meanses

beta <- c("(Intercept)" = 12, meanses = 5, cses = 2.5, sector = 1,
          "meanses:cses" = 0.7, "cses:sector" = -0.7)
s <- 6.5
su <- c(1.6, 0.3)
Su <- 0.4 * su %o% su; diag(Su) <- su^2
Lt <- chol(Su)/s
pars <- list(theta = t(Lt)[lower.tri(Lt, TRUE)], beta = beta, sigma = s)

y <- simulate(~ meanses*cses + sector*cses + (1 + cses | school),
              newparams = pars, newdata = dat)$sim_1

m <- lmer(y ~ meanses*cses + sector*cses + (1 + cses | school), dat)
fixef(m)
VarCorr(m)
\end{lstlisting}

\section{Miscellaneous}

\subsection{Intervals with maximum width}

\subsubsection{Confidence interval for $\pi$, $X \sim B(1, \pi)$}

\mytab{Context}{See \ref{sec:binomtest}. Goal is an interval for the
probability of a boy birth $\pi$ not wider than $0.02$.}

\mytab{Assumptions}
{1. Exact two-sided $(1 - \alpha)$ confidence interval\\
 2. Approximate two-sided $(1 - \alpha)$ confidence interval ($n \cdot
\bar{X} \cdot (1 - \bar{X}) > 9$)
\[
  \bar{X} \pm z_{1-\alpha/2} \, \sqrt{\frac{\bar{X} \, (1-\bar{X})}{n}}
\]}
\begin{lstlisting}
## 1.
n <- 9800                                               # sample size
x <- rbinom(1, size = n, prob = 0.5)                    # data-generating model
diff(binom.test(x, n = n, conf.level = 0.95)$conf.int)  # interval width at alpha = 0.05

## 2.
n <- 9800
x <- rbinom(1, size = n, prob = 0.5)
diff(prop.test(x, n = n, conf.level = 0.95, correct = FALSE)$conf.int)
\end{lstlisting}

\subsubsection{Confidence interval for $\mu$}

\mytab{Context}{See \ref{sec:onesamplet}. Goal is an interval for the point of
subjective equality $\mu$ not wider than 4\,Hz.}

\mytab{Assumptions}
{$X \sim N(\mu, \sigma^2)$ or $X$ arbitrary distributed for $n > 30$,\\
(approximate) two-sided $(1 - \alpha)$ confidence interval\\
1.\ $\sigma^2$ known,   $\bar{X} \pm z_{1-\alpha/2} \cdot \sigma/\sqrt{n}$\\
2.\ $\sigma^2$ unknown, $\bar{X} \pm t_{1-\alpha/2} (n-1) \cdot S/\sqrt{n}$}
\begin{lstlisting}
## 1.
n <- 55                                           
x <- rnorm(n, mean = 1000, sd = 7.5)              
diff(mean(x) + qnorm(c(0.025, 0.975))*7.5/sqrt(n))

## 2.
n <- 65
x <- rnorm(n, mean = 1000, sd = 7.5)
diff(t.test(x, conf.level = 0.95)$conf.int)
\end{lstlisting}

\subsubsection{Confidence interval for $\sigma^2$, $X \sim N(\mu, \sigma^2)$}

\mytab{Context}{See \ref{sec:onesamplet}. Goal is an interval for $\sigma^2$
not wider than (7\,Hz)$^2$.}

\mytab{Assumptions}
{Two-sided $(1 - \alpha)$ confidence interval
\[
  \left[\frac{(n - 1) \, S^2}{\chi^2_{1-\alpha/2}(n-1)},~
      \frac{(n - 1) \, S^2}{\chi^2_{\alpha/2}(n-1)}\right]
\]}
\begin{lstlisting}
n <- 55
x <- rnorm(n, mean = 1000, sd = 7.5)
diff((n - 1)*var(x)/qchisq(c(0.975, 0.025), n - 1))
\end{lstlisting}

\subsection{Multinomial processing tree (MPT) models}

\mytab{Context}{Two groups of participants will be presented with hypothetical
scenarios that require a decision whether or not to break a moral norm.
Group~1 will be instructed that breaking the norm would be beneficial to them.
The response frequencies will be analyzed using an MPT model. Goal is to
detect a reduction of Group~1 norm adherence $N_1$ of 0.2 relative to $N_2$.}

\mytab{Model}{See details in the \texttt{mpt} package \citep{WickelmaierMPT}.}

\mytab{Hypothesis}{$H_0\colon~ N_1 = N_2$}

\begin{lstlisting}
library(mpt)
spec <- mptspec("proCNI", .replicates = 2,
                .restr = list(J1 = I1, J2 = I2))   # proCNI model

n <- 400                                           # total number of subjects
m <- structure(list(                               # stub mpt object
  treeid = spec$treeid,
       n = setNames(rep(n/2, 2*8), spec$treeid),   # n/2 per group
    pcat = spec$par2prob(c(C1 = 0.4, N1 = 0.3, I1 = 0.5,   # note: order of pars
                           C2 = 0.4, N2 = 0.5, I2 = 0.5))  #   must match spec$par
), class = "mpt")

pval <- replicate(2000, {
  y <- simulate(m)
  m1 <- mpt(spec, data = y)
  m2 <- mpt(update(spec, .restr = list(N1 = N2)), data = y)
  anova(m2, m1)[["Pr(>Chi)"]][2]
})
mean(pval < 0.05)
\end{lstlisting}

\bibliographystyle{apacite}
\bibliography{$HOME/w/LMU/bibtex/long,$HOME/w/LMU/bibtex/fw}

\end{document}